\documentclass{emulateapj}
\usepackage{epsf}
\usepackage{apjfonts}
\usepackage{xspace}
\usepackage{amsmath}
\usepackage{framed} 

   \usepackage{graphicx}

\def\hide#1{}

\begin{document}

\title{The Effects of Dark Matter Annihilation on Cosmic Reionization}

\author{Alexander A.\ Kaurov\altaffilmark{1,*},  Dan Hooper\altaffilmark{2,1,3},  Nickolay Y.\ Gnedin\altaffilmark{2,1,3}}
\altaffiltext{1}{Department of Astronomy \& Astrophysics, The
  University of Chicago, Chicago, IL 60637 USA; kaurov@uchicago.edu}
\altaffiltext{2}{Particle Astrophysics Center, 
Fermi National Accelerator Laboratory, Batavia, IL 60510, USA}
\altaffiltext{3}{Kavli Institute for Cosmological Physics and Enrico
  Fermi Institute, The University of Chicago, Chicago, IL 60637 USA}
\altaffiltext{*}{kaurov@uchicago.edu} 

\begin{abstract}

We revisit the possibility of constraining the properties of dark matter (DM) by studying the epoch of cosmic reionization. Previous studies have shown that DM annihilation was unlikely to have provided a large fraction of the photons that ionized the universe, but instead played a subdominant role relative to stars and quasars. The DM, however, begins to efficiently annihilate with the formation of primordial microhalos at $z\sim100-200$, much earlier than the formation of the first stars. Therefore, if DM annihilation ionized the universe at even the percent level over the interval $z \sim 20-100$, it can leave a significant imprint on the global optical depth, $\tau$. Moreover, we show that cosmic microwave background (CMB) polarization data and future 21 cm measurements will enable us to more directly probe the DM contribution to the optical depth. In order to compute the annihilation rate throughout the epoch of reionization, we adopt the latest results from structure formation studies and explore the impact of various free parameters on our results. We show that future measurements could make it possible to place constraints on the dark matter's annihilation cross section that are at a level comparable to those obtained from the observations of dwarf galaxies, cosmic ray measurements, and studies of recombination. 

\end{abstract}

\keywords{cosmology, reionization, intergalactic medium;  FERMILAB-PUB-15-510-A}

\section{Introduction}

Despite recent advances in the observations of the high redshift universe, the physics of cosmic reionization remains uncertain. Current observations of the Ly$\alpha$ forest \citep{Becker_2015} allow us to probe the final stages of reionization, and CMB polarization data can be used to place very broad constraints on its duration \citep{Zahn_2012}.  Upcoming experiments, such as the James Webb Space Telescope (JWST), and the introduction of 21cm cosmology, will make it possible to observe the high redshift universe in much greater detail and to place much stricter constraints on the processes responsible for the reionization of the universe. 


The main objective pertaining to cosmic reionization is to determine the sources of the ionizing photons. The most widely discussed sources for hydrogen reionization are the stars within galaxies, while quasars are often thought to be primarily responsible for reionizing helium. Recent observations \citep{2015arXiv150707678M} argue in favor of quasar activity during hydrogen reionization as well. Other sources have also been proposed; for example, X-ray binaries \citep{Fialkov_2014}. 


Within this context, the annihilation of DM particles is an interesting process. The products of DM annihilation can affect the intergalactic medium (IGM), and therefore change the global ionization and thermal histories of our universe. 
%
%
In contrast to some previous studies \citep{2009PhRvD..80c5007B,2009JCAP...10..009C} we do not imagine that the DM is the only, or even the primary, source of ionizing photons (see also, \cite{2006MNRAS.369.1719M,2007MNRAS.379.1003S,2007MNRAS.375.1399R,2007MNRAS.374.1067R,2007MNRAS.377..245V,2006PhRvD..74j3502F,2008ApJ...679L..65C,2009PhRvD..80d3529N,H_tsi_2009,2010PhRvD..82j3508C}). We instead expect that the DM played a relatively minor role in cosmic reionization. However, in contrast to the stars which begin to form at $z\sim 15-20$, the DM begins to efficiently annihilate much earlier, at $z\sim100-200$. If DM ionized the universe to the level of a few percent over the redshift interval between $\sim20-200$, this would significantly impact the global optical depth, $\tau$. We will show that the constraints on $\tau$ from the combination of future CMB and 21cm observations will provide a powerful tool for constraining the properties of particle DM.

The remainder of this paper is structured as followed. 
First, in \S\ref{sec:boosting}, we evaluate the total rate of DM annihilation, including the boost factor, which quantifies the amount of structure over cosmic history. In \S\ref{sec:DMspectrum}, we discuss in the interactions between the DM annihilation products and the IGM. Finally, we present in \S\ref{sec:results} current and projected constraints on the dark matter annihilation cross section. We summarize these constraints and discuss the prospects for the further developments in \S\ref{sec:discussion}.

\section{Boost factor}
\label{sec:boosting}

The global rate of DM annihilation at a given redshift is proportional to $\langle n_{DM}^2 \rangle$, where $n_{DM}$ is the number density of DM particles. As a result of inhomogeneities in the DM density, the annihilation rate is enhanced by the following ``boost factor'':
\begin{equation}
B = \dfrac{\langle n_{DM}^2 \rangle}{\langle n_{DM} \rangle ^2}.
\end{equation}

In order to calculate the boost factor properly, one needs to integrate over the relevant volume, and over all scales. Unfortunately, simulations of large scale structure do not have the resolution required to characterize such structure on very small scales. Therefore, as in previous studies (\citet{Belikov_2009,H_tsi_2009,2009JCAP...10..009C}), we must rely on extrapolations of the halo mass function and the relationship between halo mass and concentration.


We attempt to approach this problem systematically, by parameterizing the uncertainties regarding the distribution of dark matter and then classifying the possible values for the parameters into three categories, which we label as \textit{Low}, \textit{Medium}, and \textit{High} (see Table 1). When we consider this full range of parameters, the boost factors calculated during reionization span approximately two orders of magnitude; see Figure \ref{fig:extremes}. In the following subsections, we will discuss each of these individual parameters and their impact on the DM annihilation rate.

\begin{center}
\begin{table}
\caption{Free parameters and their adopted values.}
    \begin{tabular}{ c c c c }
    \textbf{Parameter description}  & \textbf{Low} & \textbf{Medium} & \textbf{High} \\
    \S\ref{subsec:mf} Mass function cut-off (in $\log_{10}M_\odot$)        & -6 & -9 & -12 \\
    \S\ref{subsec:nfw} Modified NFW profile for small halos & No & Yes &  Yes \\
    \S\ref{subsec:conc} Scatter of concentrations (in $\sigma_{\log_{10}c}$)              & 0.08 & 0.16 & 0.24 \\
    \S\ref{subsec:subhalos} Subhalo mass function & No & Yes & Yes \\
    \S\ref{subsec:nonspherical} Caustics and non-spherical profiles & None & None & 2.0 \\ \\
   \end{tabular}
\end{table}
\end{center}

\subsection{Halo Mass Function}
\label{subsec:mf}

Halo mass function models can be based on analytic calculations, or fit to the results of numerical simulations. Most models based on numerical simulations (for example, \citet{Tinker_2008} and its extension to higher redshifts by \citet{Behroozi_2013}) are tuned to match the characteristics of large halos, $10^{8}M_\odot < M < 10^{15}M_\odot$. Lower mass halos, which we are particularly interested in, can also be explored numerically, but require dedicated simulations, such as those carried out by \citet{Diemand_2005}. 
%
Studies such as these find a behavior of $dN/dM \propto M^{-2}$, which can be modeled with a simple Press-Schecter mass function \citep{Press_1974}. Although there are known deviations from the Press-Schecter halo mass function, these are important only at redshifts below $\sim 20$ and for masses above $\sim10^6-10^8 M_{\odot}$. In our calculations, we neglect halos more massive than $10^6 M_\odot$, for which the baryonic content increases the rate of atomic processes that affect the local IGM, effectively reducing the contribution to global reionization \citep{Kaurov2015}; this choice has little impact on our results.

\begin{figure}
\begin{center}
\includegraphics[width=1\columnwidth]{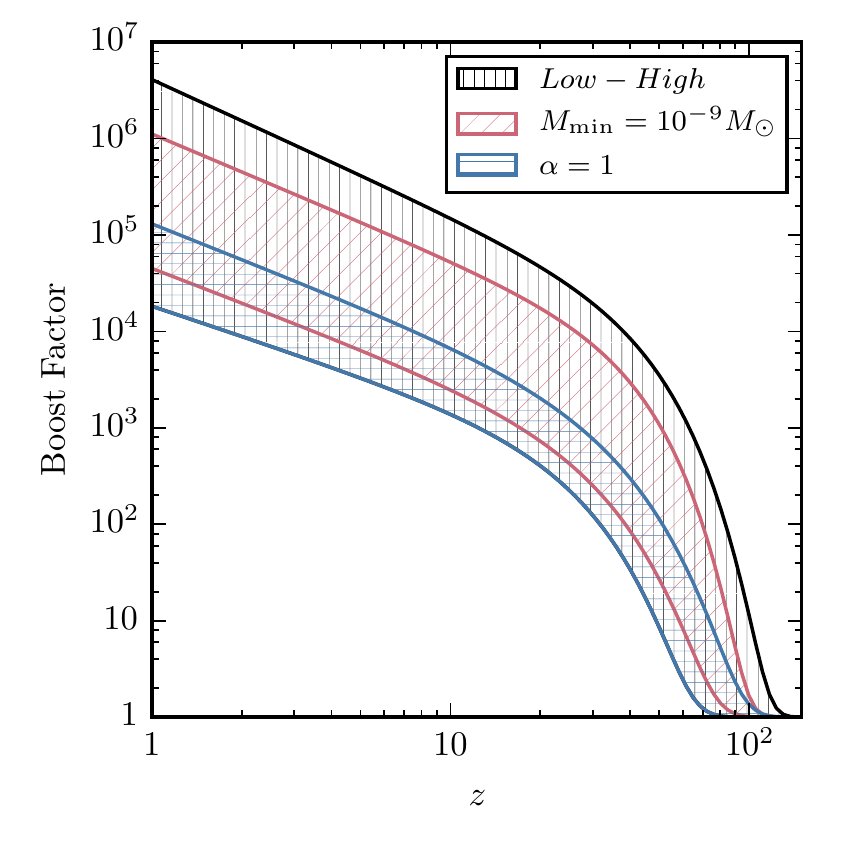}
\caption{\label{fig:extremes}
The range of boost factors evaluated using the \textit{Low} and \textit{High} parameters listed in Table 1. The black and vertically hatched region spans the entire range of these parameters. The red and diagonally hatched region is the same, but with $M_\mathrm{min}$ fixed to $10^{-9}M_\odot$. The blue and horizontally hatched region has the value of the inner slope, $\alpha$, fixed to unity.}
\end{center}
\end{figure}

The halo mass function is further predicted to be truncated below a minimum mass, $M_{\rm min}$, which is determined by the temperature at which the DM became kinetically decoupled. Although predictions for $M_{\rm min}$ depend on the specific interactions between the DM and the Standard Model, and are thus highly model dependent, most models of DM as weakly interacting massive particles (WIMPs) feature values in the range of $10^{-3}M_\odot$ to $10^{-12}M_\odot$~\citep{Green_2004, Profumo_2006}.


In Figure \ref{fig:extremes}, we plot the range of boost factors found when allowing $M_\mathrm{min}$ to vary between $10^{-6}M_{\odot}$ and $10^{-12}M_{\odot}$ (black, vertically hatched), and when fixed to $10^{-9} M_{\odot}$ (red, diagonally hatched). The minimum mass is the second most important parameter we have considered, after the inner slope of the halo profile which we discuss in the next subsection.

\subsection{Halo Profiles}
\label{subsec:nfw}

Beginning with the \textit{Low} case, we adopt the standard Navarro-Frenk-White (NFW) radial profile for DM halos \citep{1997ApJ...490..493N}. Other types of commonly adopted profiles ({\it e.g.} Einasto) do not significantly modify our results, as most of the DM annihilation takes place around the scale radius, where such profiles are very similar.  An exception, however, can be found for profiles with much steeper inner slopes. It is common to generalize the NFW profile such that its inner slope, $\alpha$, is treated as a free parameter:
\begin{equation}
\rho(r) = \dfrac{\rho_0}{(r/r_S)^\alpha(1+r/r_S)^{(3-\alpha)}}.
\end{equation}
For the standard NFW case, $\alpha\equiv 1$. For the \textit{Medium} and \textit{High} cases, we adopt the fit for $\alpha$ provided by \citet{Ishiyama_2014}:
\begin{equation}
\label{eq:alpha}
\alpha = - 0.123 \log(M_{\rm vir}/M_\mathrm{min}) + 1.461,
\end{equation}
where $M_{\rm vir}$ is the virial mass of a halo and $M_\mathrm{min}$ is the minimum halo mass. For masses that yield a value less than unity, we adopt $\alpha=1$. This parameterization significantly increases the annihilation rate in the smallest halos relative to the standard NFW prescription.

In Figure \ref{fig:extremes} we show the boost factor calculated in the $Low$ and $High$ cases, with $\alpha$ fixed to unity (blue, horizontally hatched). This illustrates that the inner profile of the smallest mass halos can significantly impact the global boost factor. We note that although such steep profiles for the smallest halos are supported by simulations \citep{Ishiyama_2014}, it is not yet clear whether this behavior has been reliably resolved, and will require further studies to confirm.

\begin{figure}
\begin{center}
\includegraphics[width=1\columnwidth]{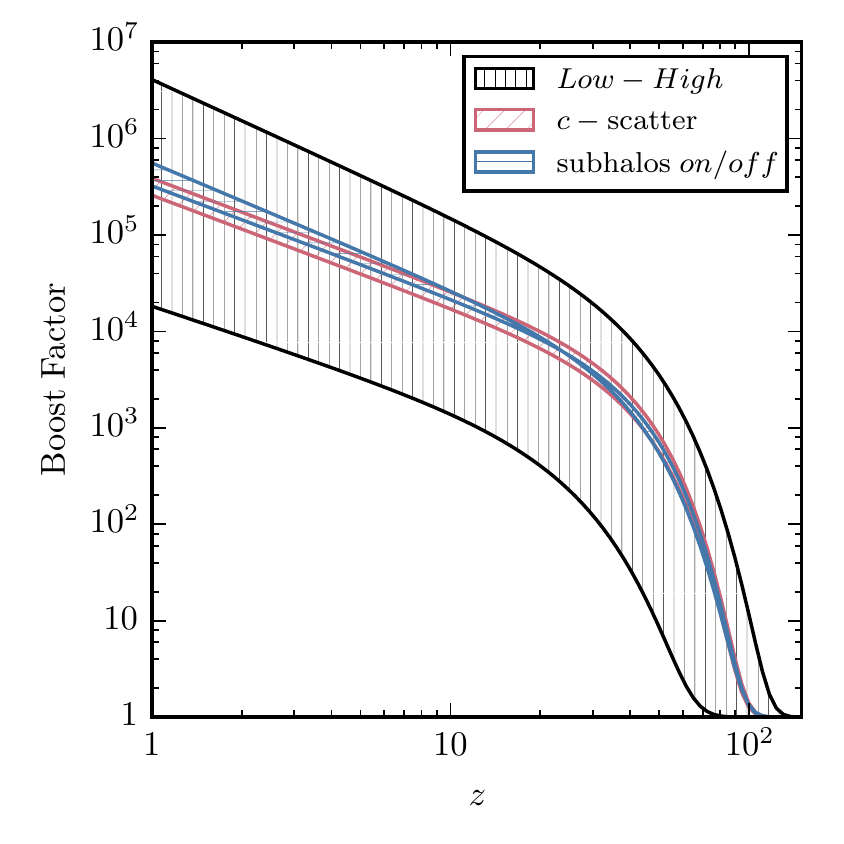}
\caption{\label{fig:subhalos}
The range of boost factors associated with variations in the distribution of halo concentrations (red, diagonally hatched) and with variations in the treatment of subhalos (blue, horizontally hatched), while keeping all other parameters fixed to their \textit{Medium} values (see Table~1).  The black hatched region denotes the full range of boost factors between the \textit{Low} and \textit{High} parameter sets.}
\end{center}
\end{figure}

\subsection{Halo Concentrations}
\label{subsec:conc}

For the concentrations of our DM halos, we adopt the model of \citet{Diemer_2015}. Again, since this model is based on simulations, only halos with masses greater than $\sim10^{10}\;M_\odot$ are directly probed. However, given that these concentrations are in fairly good agreement with those from simulations of high-redshift ($z\sim 30$) microhalos (see Figure 8 of \citet{Diemer_2015}), we are optimistic about the reliability of this application. 


Even though the average concentration for a halo of a given mass is well defined, there is a significant degree of halo-to-halo variation in this quantity. In \citet{Diemer_2015}, the authors report that the 68\% rms scatter in $\log_{10}c_{200c}$ is $\sim 0.16$ around the median value. Depending on the underlying distribution, such variations can increase the boost factor by $15-50\%$. We adopt the probability distribution function for concentrations as described in \cite{2014arXiv1412.4308M}. For our \textit{Low}, \textit{Medium} and \textit{High} scenarios, we adopt values of $\sigma_{\log_{10} c}=$ 0.08, 0.16 and 0.24, respectively. The impact of this variation is small compared to other parameters considered in this study (see Figure \ref{fig:subhalos}).

\subsection{Subhalos}
\label{subsec:subhalos}

The halo mass function adopted in this study does not account for the subhalos that reside within larger halos. The presence of such subhalos is predicted to enhance the DM annihilation rate. To estimate their impact, we followed the approach of \citet{Sanchez_Conde_2014}, finding that the presence of subhalos does not increase the global boost factor by a large factor (see Figure \ref{fig:subhalos}). More specifically, while subhalos can play a significant role in the largest halos ({\it e.g.} \citet{2015arXiv150802713Z}), such very massive halos are rare, especially at $z>10$. For small halos at high redshifts, the impact of subhalos is insignificant due to the flattening of the concentration function at low masses.  Since the presence of subhalos does not strongly impact our results, we do not explore additional effects, such as the dependence of the subhalo mass function on the concentration of the host halo \citep{Mao_2015, Emberson_2015}.

\subsection{Non-Spherical Halos}
\label{subsec:nonspherical}

Thus far in this study, we have assumed that DM halos and subhalos are spherically symmetric. Departures from sphericity can, however, be present during the active collapse of the first halos and during the subsequent mergers of halos. This can increase the boost factor; for example, \citet{Anderhalden_2013} found that departures from sphericity enhance in the annihilation rate by a factor of $\sim 1.5$ at $z \sim 30$ for primordial micro-halos, and by an additional factor of $\sim 1.5$ due to caustics. Motivated by these results, we double the total boost factor in our \textit{High} model.


\begin{figure}[h!]
\begin{center}
\includegraphics[width=1\columnwidth]{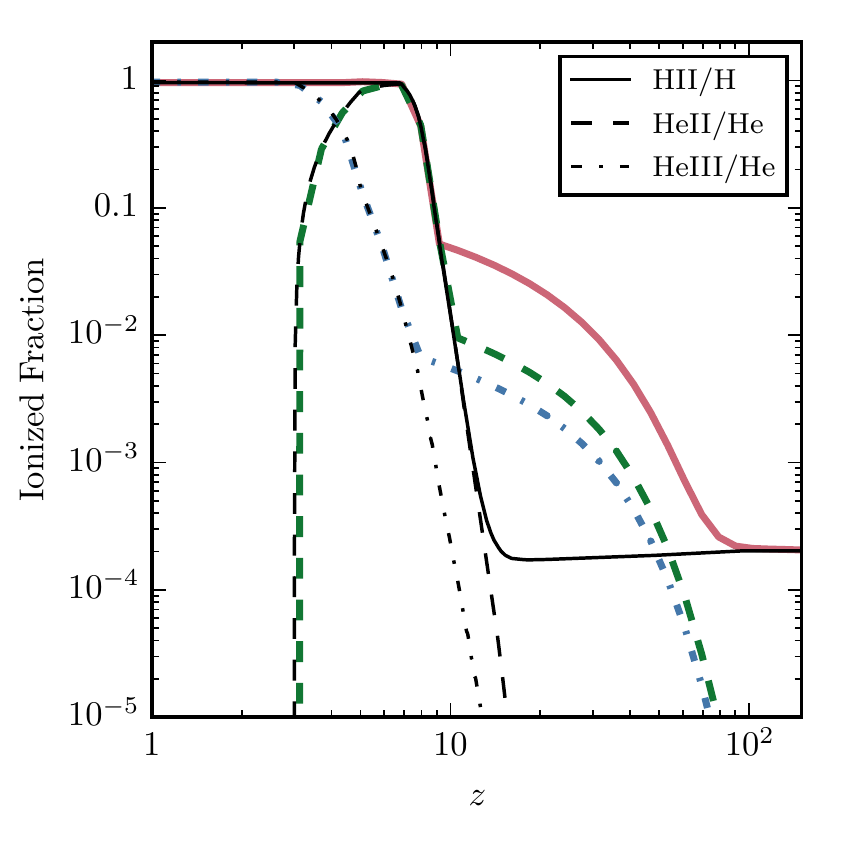}
\caption{\label{fig:rhist}
The fraction of ionized hydrogen (solid), helium II (dashed) and helium III (dot-dashed) as a function of redshift in our fiducial model.  The colored lines include the contribution from a 40 GeV DM particle that annihilates to $b\bar{b}$ with a cross section of $\sigma v = 10^{-26}$ cm$^3$/s, adopting the \textit{High} boost factor model (see Table I). In this particular case, the effects of DM annihilation increase the optical depth, $\tau$, from 0.059 to 0.068.}
\end{center}
\end{figure}

\section{Cosmic Ionization By DM Annihilation}
\label{sec:DMspectrum}

\subsection{Fiducial Reionization Model}

In Figure \ref{fig:rhist}, we plot our fiducial reionizatoin model, which was derived from one of the numerical simulations completed as part of Cosmic Reionization on Computers (CROC) project \citep{Gnedin_2014, Gnedin_Kaurov_2014}. The optical depth of this fiducial model is $\tau=$0.059, compared to the value of $\tau=0.066\pm0.016$ derived from polarization and temperature measurements by the \citet{2015arXiv150201589P}. 
In realistic models (those that are not ruled out by other indirect detection probes), DM annihilation can ionize the universe to the level of a few percent or less before stars begin to form. Therefore, we can safely assume that the propagation of ionization fronts is not strongly affected by this small uniform ionized fraction.

\subsection{Efficiency of DM Annihilation in Ionizing the IGM}
\label{subsec:lh}

The annihilation of DM is characterized by the mass of the DM particle, the annihilation cross section, $\langle\sigma v \rangle$, and the products of those annihilations. The effects of DM annihilation on the IGM depend strongly on the species and energies of the particles that are created in this process. We use the results of \citet{Cirelli_2011} and \citet{Ciafaloni2011} to account for the hadronization and cascades of the DM annihilation products.\footnote{http://www.marcocirelli.net/PPPC4DMID.html}


Relatively low energy photons and electrons are the most efficient in reionizing the IGM. This is in contrast to the effects of DM annihilation on recombination, for which high energy electrons are also very effective \citep{Kaurov2015,Shull_1979,Shull_1985,Dalgarno_1999,Furlanetto_2010,Vald_s_2010,2013PhRvD..87j3522D,2015arXiv150603812S}. The main difference between these cases is the inhomogeneity of the baryon distribution. In \citet{Kaurov2015}, it was shown that overdensities in baryons effectively reduce the production of energetic photons and increase the rate of atomic processes that affect the IGM only locally. Therefore, halos containing baryons are generally only able to ionize themselves.

We roughly estimate that the transition between halos with and without baryons occurs around $\sim 10^6 M_\odot$, as defined by the filtering mass~\citep{2013ApJ...763...27N}. The contribution to reionization from DM annihilation in larger halos is expected to be suppressed. We also note that once halos with baryons begin to appear, star formation begins and quickly overtakes DM annihilation as a source of ionizing photons.

\subsection{Ionization Equilibrium During the Dark Ages}

In order to estimate the global ionization fraction during the dark ages, we assume spatially uniform ionization by DM annihilation. Even though the large boost factors indicate that most of the DM annihilation takes place within halos, these halos are distributed relatively uniformly due to the flat bias function at low masses. Furthermore, the radiation produced through DM annihilation typically has a long mean free path, exceeding the average distance between primordial halos.

The process of ionization competes with the recombination of ionized particles with electrons. The global recombination rate of hydrogen is given by (a similar equation can be written for helium ions):
\begin{equation}
R = C \bar{n}_{HII} \bar{n}_e \alpha_H,
\end{equation}
where $\alpha_H$ is a recombination coefficient, $\bar{n}_{HII}$ and $\bar{n}_e$ are number densities of protons and electrons, and $C$ is the clumping factor, which characterizes the substructure of baryons (analogous to the boost factor for DM annihilation). At high redshifts, when little baryonic structure exists, the clumping factor is of order unity. 

From the rate of ionization and recombination at each redshift, we can calculate the abundance of each ion. Although the small ionized fraction should affect the propagation of ionization fronts created by galaxies, we   neglect this affect and assume that our fiducial reionization model is not altered dramatically below redshift $\sim10$ (as is supported, for example, in Figure \ref{fig:rhist}).

\hide{TODO: Include citations 
\cite{Slatyer_2009}
\cite{H_tsi_2009}
\cite{Evoli_2013}}

\begin{figure*}[h!]
\begin{center}
\includegraphics[width=1\columnwidth]{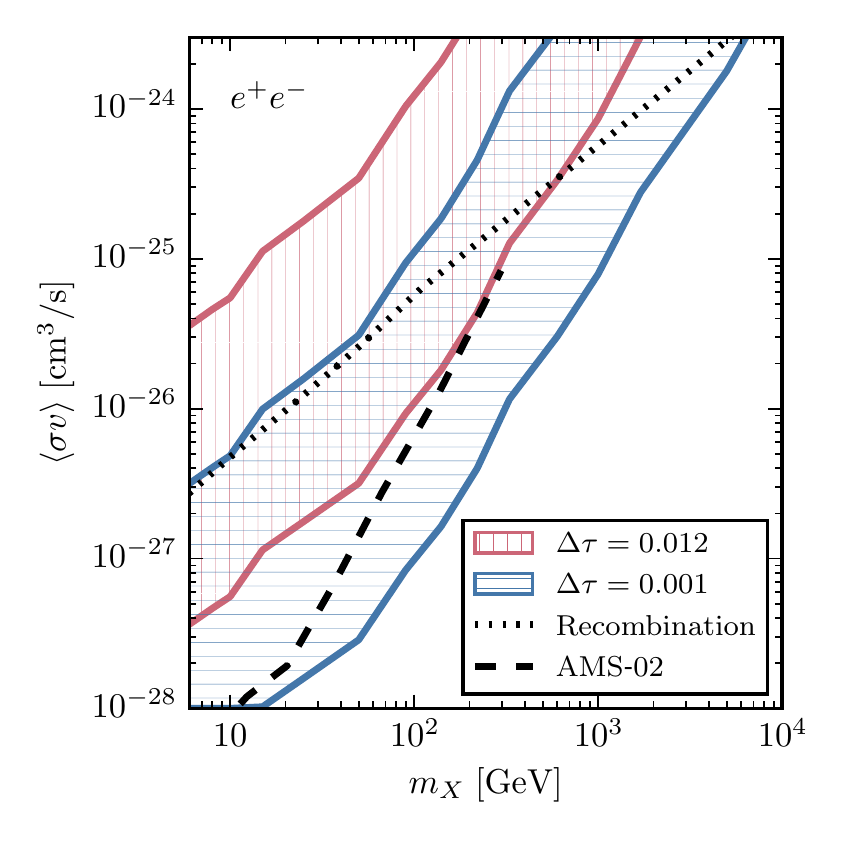}
\includegraphics[width=1\columnwidth]{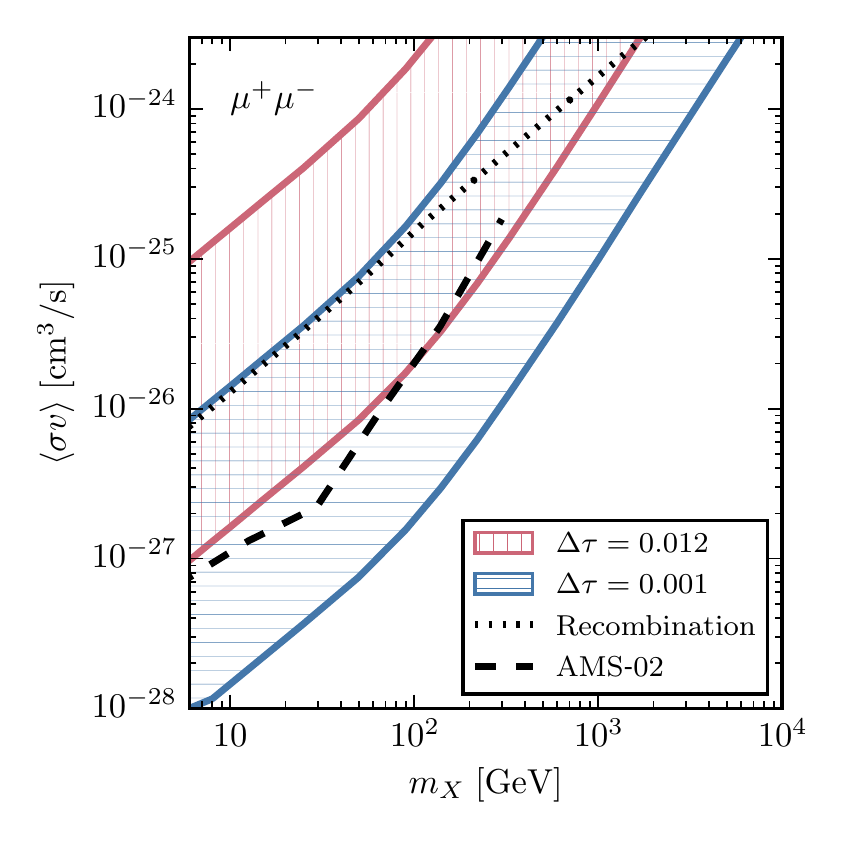}
\includegraphics[width=1\columnwidth]{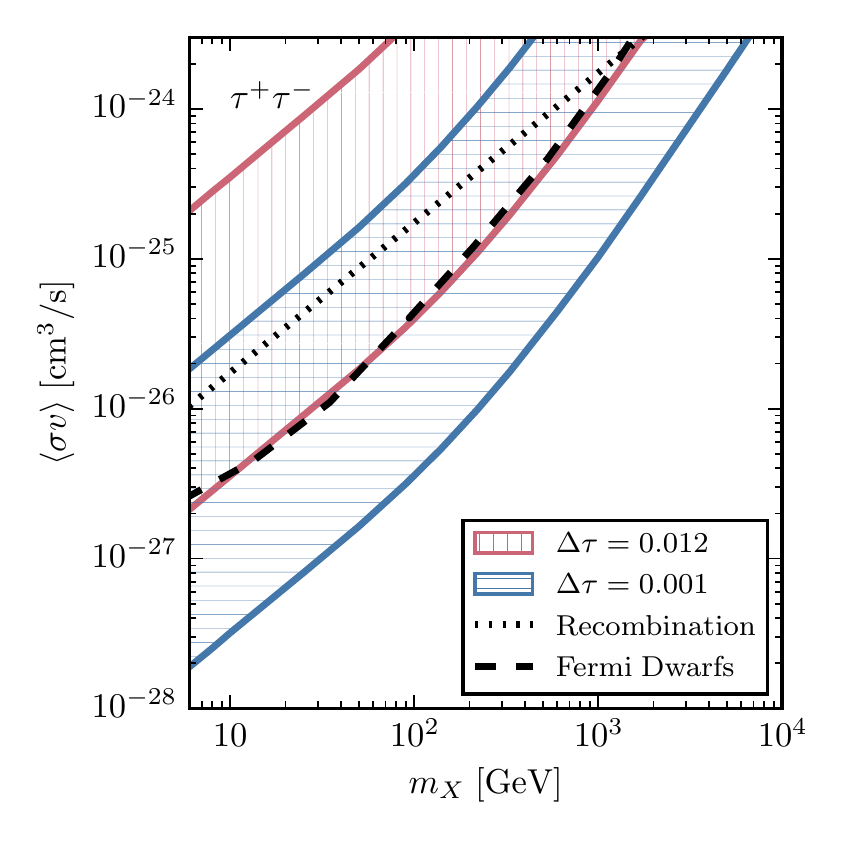}
\includegraphics[width=1\columnwidth]{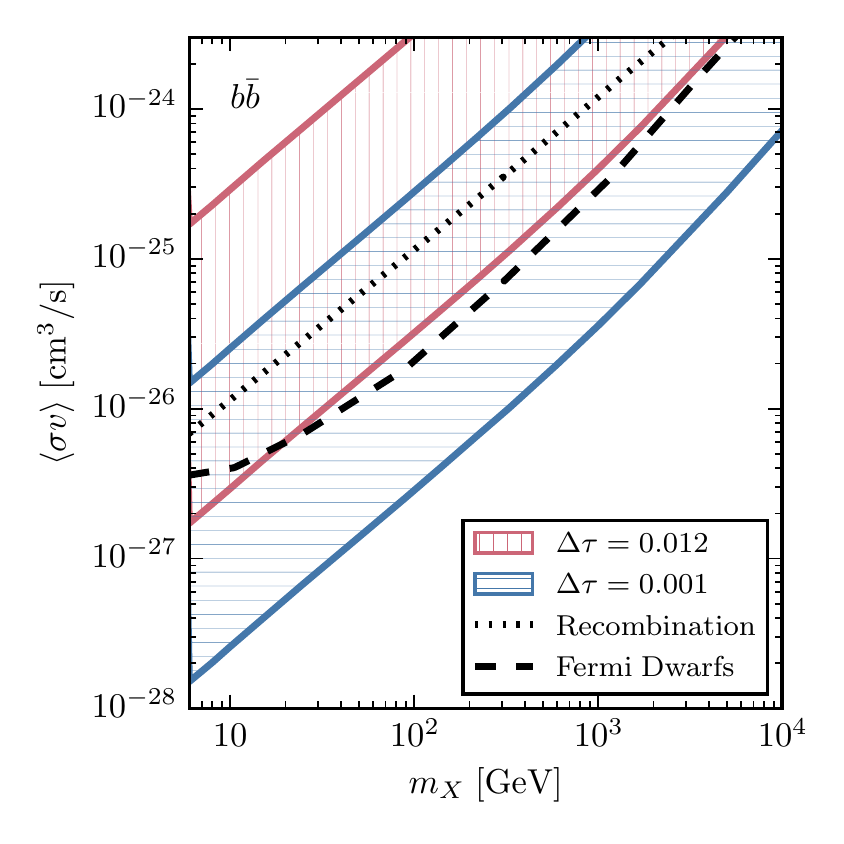}
\caption{\label{fig:sigmavmx}
Constraints on the DM annihilation cross section for $e^+ e^-$, $\mu^+ \mu^-$, $\tau^+\tau^-$ and $b\bar{b}$ final states. The red vertically hatched regions represent the constraints from current measurements, corresponding to $\Delta\tau=$0.012, while the blue horizontally hatched regions are the constraints projected from future 21 cm measurements with $\Delta \tau =0.001$. The width of these regions reflect the range of \textit{Low} to \textit{High} boost factor models (see Table I). For comparison, we show the constraints from the impact of DM annihilation on recombination \citep{2015arXiv150201589P,Slatyer_2009} (dotted), as well as from gamma-ray observations of dwarf galaxies~\citep{2015arXiv150302641F} and measurements of the cosmic ray positron spectrum~\citep{2013PhRvL.111q1101B} (dashed).}
\end{center}
\end{figure*}

\begin{figure*}
\begin{center}
\includegraphics[width=1\columnwidth]{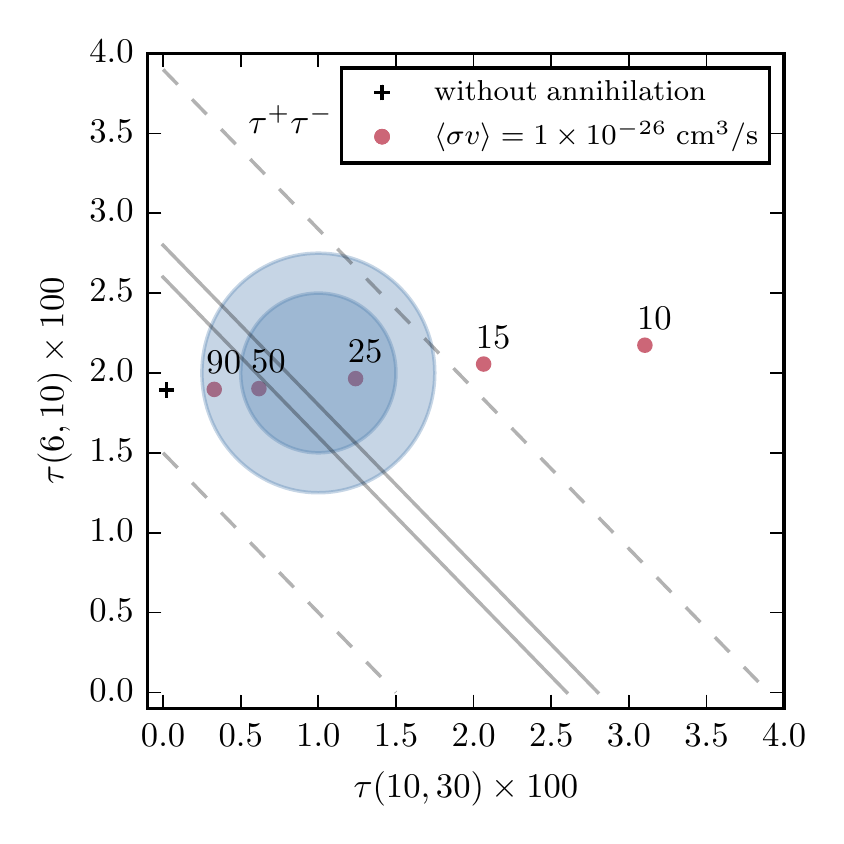}
\includegraphics[width=1\columnwidth]{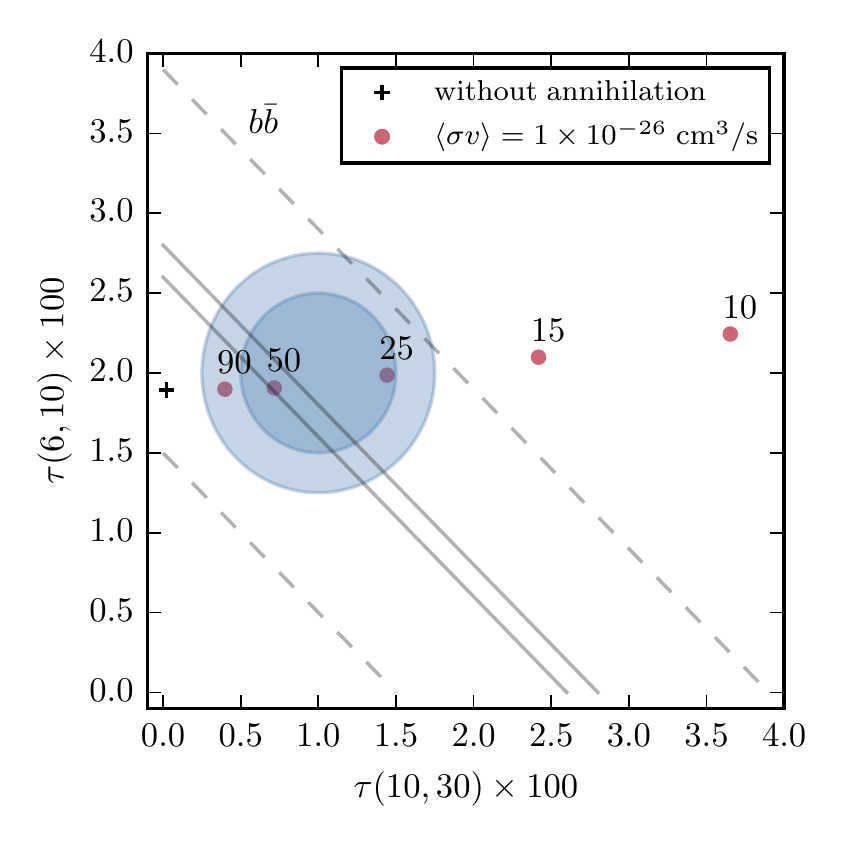}
\caption{\label{fig:taudiff}
Constraints on the optical depth integrated over two redshift ranges: $z=6-10$ and $z=10-30$. The dashed and solid gray lines are contours of constant total optical depth, $\tau=0.066\pm0.012$ (current precision, \citet{2015arXiv150201589P}) and $\tau=0.066\pm0.001$ (as expected with future 21 cm data, \citet{2015arXiv150908463L}), respectively. The blue shaded regions are the estimated constraints from a principal component analysis of cosmic variance limited CMB polarization data \citep{2007ApJ...657....1M}. The black cross denotes our fiducial reionization model, without any contribution from DM annihilation, while the red points denote the predictions including DM annihilating with a cross section $\langle\sigma v \rangle = 1\times10^{-26}\,\mathrm{cm^2}$, for various values of the DM mass (in GeV) and for our \textit{High} boost factor model. Results are shown for DM annihilating to $\tau^+ \tau^-$ (left) and $b\bar{b}$ (right).}
\end{center}
\end{figure*}

\section{Results}
\label{sec:results}

For a given particle DM model and model for the redshift-dependent boost factor, we can calculate the effects of DM annihilation on the IGM and determine the evolution of the universe's ionization fraction. We show such an example in Figure \ref{fig:rhist}, for the representative example of a 40 GeV DM particle that annihilates to $b\bar{b}$ with a cross section of $\sigma v = 10^{-26}$ cm$^3$/s, adopting the \textit{High} boost factor model. In this model, the ionization fraction at $z \sim 10-50$ increases to the level of a few percent. As a result, the total optical depth is enhanced significantly, from 0.059 to 0.068. This illustrates that precision measurements of $\tau$ could potentially enable us to place interesting constraints on the properties of the DM.

Polarization measurements by Planck \citep{2015arXiv150201589P} using the Low Frequency Instrument at large angular scales, combined with Planck temperature and lensing data, yield a total reionization optical depth of $\tau=0.066\pm0.016$. If these measurements are combined with those of baryon acoustic oscillations, Type Ia supernovae, and the local Hubble constant (see Table 4 of \citet{2015arXiv150201589P}), the error on this quantity is further reduced to $\Delta\tau=0.012$. Upcoming 21 cm measurements will be a very powerful tool for further constraining the contribution to the optical depth from stars \citep{2015arXiv150908463L}, and is expected to enable us to reduce the uncertainty on this parameter to the level of $\Delta\tau\sim0.001$.

In Figure \ref{fig:sigmavmx}, we plot our constraints on the DM annihilation cross section, for the current measurement uncertainty of $\Delta\tau=0.012$ (red, vertically hatched), and for a future measurement with $\Delta \tau = 0.001$ (blue, horizontally hatched). Each of these results is shown as a band, which covers the range of boost factor models from \textit{Low} to \textit{High}, as described earlier in this paper. These results are compared to the constraints derived from observations of dwarf spheroidal galaxies by the Fermi Gamma-Ray Space Telescope~\citep{2015arXiv150302641F}, measurements of the cosmic-ray positron spectrum by AMS-02~\citep{2013PhRvL.111q1101B}, and from the impact of DM annihilation on recombination \citep{2015arXiv150201589P,Slatyer_2009}. For high-mass DM particles, our constraints are less competitive with those from other observations, as high-energy electrons and photons do not interact significantly with the IGM and thus lead to very inefficient reionization.


The global optical depth, $\tau$, receives contributions from stars, quasars, and (possibly) annihilating DM. Even with a very high-precision measurement of $\tau$, it may still be difficult to distinguish between these contributions. This quantity is not, however, the only relevant information contained in the CMB. Following the method presented by \cite{2007ApJ...657....1M}, one can decompose $\tau$ into redshift bins, allowing us to separate the early effects of DM annihilation from the later effects of stars and quasars.  With this goal in mind, we adopt redshift bins of $6-10$ and $10-30$, and plot these results in Figure \ref{fig:taudiff}, along with existing and projected constraints from CMB polarization measurements. We show results for DM annihilating to $\tau^+ \tau^-$ (left) and $b\bar{b}$ (right), with a cross section of $10^{-26}$ cm$^3$/s. Each numbered red point represents the prediction for a DM particle of a given mass, adopting our \textit{High} boost factor model. Such a decomposition could plausibly be used to distinguish the effects of DM annihilation from astrophysical sources of reionization, and perhaps even provide an approximate measurement of the DM particle's mass.

\section{Discussion and Summary}
\label{sec:discussion}

Given the current state of observation and theory, it is not yet possible to use the reionization history of the universe to place strong constraints on annihilating DM. There are compelling reasons, however, to expect that this may change in the coming years. Uncertainties regarding the profiles, concentrations, and other features of low-mass DM halos and subhalos are likely to be reduced as simulations improve. In parallel, improvements in hydrodynamical simulations, combined with empirical input from JWST and 21 cm observations, will enable us to more accurately predict the contribution to reionization from stars and quasars. Finally, determinations of the universe's optical depth, $\tau$, are expected to become much more accurate as CMB polarization and 21 cm measurements proceed. Ultimately, the universe's optical depth could be decomposed into redshift bins, allowing us to separate the early effects of DM annihilation from lower redshift sources of ionizing photons. Taken together, it appears plausible that the reionization history of the universe could, in the foreseeable future, provide a valuable and complementary probe of annihilating DM, allowing us to place constraints on the DM's mass, annihilation cross section and channel, and even the minimum halo mass, as determined by the temperature of kinetic decoupling.

Lastly, we note that the heat produced through DM annihilation could also impact the evolution of the IGM. The temperature of the IGM and CMB decouple at redshift $z \sim140$, after which the gas cools more quickly than radiation. Later, this gas is reheated during reionization by stars, although there are proposed mechanisms that could pre-heat the IGM prior to this stage, including X-rays from high-mass X-ray binaries \citep{Jeon_2014} and supermassive black holes \citep{Tanaka_2012}. The rising temperature of the IGM increases the filtering scale, reducing the clumping factor \citep{Jeon_2014}, thus decreasing the recombination rate and speeding reionization. The complexity and interconnection of these effects can be reliably studied only in numerical simulations.

\acknowledgements
Fermilab is operated by Fermi Research Alliance, LLC, under Contract No. DE-AC02-07CH11359 with the United States Department of Energy. This work was also supported in part by the NSF grant AST-1211190.

\bibliographystyle{apj}
\bibliography{refs,ng-bibs/igm,ng-bibs/self,ng-bibs/dsh,ng-bibs/rei,ng-bibs/21cm,ng-bibs/newrei}

\end{document}